\documentclass[aps,pre,showpacs,amsmath,amsfonts,amssymb,superscriptaddress]{revtex4}

\newcommand{\beq}{\begin{equation}}
\newcommand{\eeq}{\end{equation}}
\newcommand{\bea}{\begin{eqnarray}}
\newcommand{\eea}{\end{eqnarray}}

\newcommand{\f}{\begin{equation}}
\newcommand{\ff}{\end{equation}}

\input{psfig.sty}

\begin{document}


\title{Topology and phase transitions: from an exactly solvable model to a
relation between topology and thermodynamics}

\author{Luca Angelani}
\email{luca.angelani@phys.uniroma1.it}
\affiliation{Dipartimento di Fisica and INFM - CRS SMC, 
Universit\`a di Roma {\em La Sapienza}, P.le A. Moro 2, I-00185 Roma, Italy}

\author{Lapo Casetti}
\email{casetti@fi.infn.it}
\altaffiliation[also at: ]
{Centro Interdipartimentale per lo Studio delle Dinamiche Complesse (CSDC),
Universit\`a di Firenze, Istituto Nazionale per la Fisica della Materia (INFM),
Unit\`a di Ricerca di Firenze, and Istituto Nazionale di Fisica Nucleare
(INFN), Sezione di Firenze, via G.~Sansone 1, I-50019 Sesto Fiorentino (FI),
Italy} 
\affiliation{Dipartimento di Fisica, Universit\`a di Firenze, 
via G.\ Sansone 1,
I-50019 Sesto Fiorentino (FI), Italy} 

\author{Marco Pettini}
\email{pettini@arcetri.astro.it}
\altaffiliation[also at: ]
{Centro Interdipartimentale per lo Studio delle Dinamiche Complesse (CSDC),
Universit\`a di Firenze, Istituto Nazionale per la Fisica della Materia (INFM),
Unit\`a di Ricerca di Firenze, and Istituto Nazionale di Fisica Nucleare
(INFN), Sezione di Firenze, via G.~Sansone 1, I-50019 Sesto Fiorentino (FI),
Italy} 
\affiliation{Istituto Nazionale di Astrofisica (INAF),
Osservatorio Astrofisico di Arcetri, Largo E.~Fermi 5,
I-50125 Firenze, Italy} 

\author{Giancarlo Ruocco}
\email{giancarlo.ruocco@roma1.infn.it}
\affiliation{Dipartimento di Fisica and INFM - CRS Soft, 
Universit\`a di Roma {\em La Sapienza}, P.le A. Moro 2, I-00185 Roma, Italy}

\author{Francesco Zamponi}
\email{francesco.zamponi@phys.uniroma1.infn.it}
\affiliation{Dipartimento di Fisica and INFM, 
Universit\`a di Roma {\em La Sapienza}, P.le A. Moro 2, I-00185 Roma, Italy}

\begin{abstract}
The elsewhere surmised topological origin of phase
transitions is given here new important evidence through the analytic study
of an exactly solvable model for which both topology and thermodynamics are
worked out. The model is a mean-field one with a $k$-body interaction. It
undergoes a second order phase transition for $k=2$ and a first order one 
for $k>2$. This opens a completely new perspective for the understanding of 
the deep origin of first and second order phase transitions, respectively.
In particular, a remarkable theoretical result consists of a new mathematical
characterization of first order transitions. Moreover, we show that a
``reduced'' configuration space can be defined in terms of collective variables,
such that the correspondence between phase transitions and topology changes
becomes one-to-one, for this model. Finally,
an unusual relationship is worked out between the microscopic description of 
a classical $N$-body system and its macroscopic thermodynamic behaviour. This
consists of a functional dependence of thermodynamic entropy upon the Morse
indexes of the critical points (saddles) of the constant energy
hypersurfaces of the microscopic $2N$-dimensional phase space. Thus 
phase space (and configuration space) topology is directly related to 
thermodynamics.

\end{abstract} 

\date{June 30, 2004}


\maketitle

\section{Introduction}

Thermodynamical phase transitions are certainly one of the main topics of
statistical physics.
A huge amount of work, both theoretical and experimental, has been
done during the past decades leading to remarkable successes as is witnessed, 
for example,  by the Renormalization Group theory of critical 
phenomena. However, there are still longstanding open problems about
phase transitions; among them we can mention amorphous and disordered systems
(like glasses and spin-glasses) undergoing ``dynamical'' transitions, or  
first-order phase transitions, which are still lacking a satisfactory 
theoretical understanding of their origin. 
Moreover, on the forefront of modern research in statistical physics, 
standard theoretical definitions and methods are
challenged by the experimentally observed phase transitions occurring  in 
small classical and quantum systems (nano and mesoscopic) like atomic and
molecular clusters, polymers and proteins, Bose-Einstein condensates, 
droplets of quantum liquids, etc.
Finally, in the mathematically rigorous background of phase transitions 
theory, neither in the Yang-Lee theory for the grandcanonical ensemble 
\cite{YL} nor in the Ruelle, Sinai, Pirogov theory for the canonical ensemble 
\cite{RS}, an \emph{a-priori} mathematical distinction can be made among the 
potentials leading to first or second order phase transitions, respectively.

The present paper aims to contribute to the advancement of a recently 
proposed theoretical framework where the singular behaviours of thermodynamic 
observables at a phase transition are attributed to major topology changes 
in phase space and -- equivalently -- in configuration space
\cite{cccp,franzosi,phi4,physrep,xy}. 
More precisely, 
in Ref. \cite{cccp,franzosi,phi4,physrep,xy},
it has been
proposed that thermodynamic phase transitions could be a consequence of
suitable topology changes of certain submanifolds of the configuration space
defined by the potential energy function. The presence of such topological
changes has been recently shown to be a \emph{necessary} condition, under fairly
general assumptions, for the
presence of a phase transition \cite{theorem}; however, the converse is
\emph{not} true, and no rigorous results are available yet on the sufficient
conditions. The analytical or numerical study of particular models thus remains
crucial to get hints towards more general results (see 
also Refs. \ \cite{grinza,kastner} for recent results on one-dimensional 
systems, and Ref.\ \cite{ribeiro} as to the fully connected spherical model). 

In this perspective, we stress that 
the topological approach has been hitherto applied only to systems undergoing 
second-order
phase transitions.
Therefore tackling first order phase transitions in the topological framework
is of great interest, because some new insight into the challenging 
problem of their origin can be obtained, and because this reinforces the
working hypothesis that  the topological approach could unify the treatment 
of the different kinds of phase transitions in view of encompassing also more 
``exotic'' ones, like glassy transitions and the others mentioned above.

In this paper we first study a model which, according to the value of a parameter,
has no transition or undergoes a first or a second order transition. 
Remarkably, for this model an exact analytical computation is possible of the 
Euler characteristic -- that is a topological invariant -- of those 
submanifolds of the configuration space whose topological changes are expected
to be related to the phase transition. 
We find \cite{epl2003} that the phase transition is actually signaled by a discontinuity 
in the Euler characteristic $\chi(e)$ and that the sign of the second
derivative of $\chi(e)$ indicates the order of the transition.
Then, we study the topology of
submanifolds of a ``reduced'' configuration space, i.e., the space of some
collective variables: in this case we find a one-to-one correspondence between
phase transitions and topology changes.

Finally, we derive a general result
showing that an analytic {\it estimate} 
of another topological invariant of the same submanifolds can be worked out
which allows to directly link topology and thermodynamic entropy.

\section{A key study}

In this Section we present a study of 
the thermodynamical properties of the mean-field 
$k$-Trigonometric Model ($k$TM), as well as of the topological properties of its
configuration space. A preliminary study of this model along these lines 
has already been reported
in \cite{epl2003}; there only the microcanonical thermodynamics was considered,
while here we are going to discuss also the canonical thermodynamical
properties. Being a model with long-range interactions which may undergo also
first-order phase transitions, we expect canonical and microcanonical
thermodynamic functions to be different, at least close to first-order
transitions \cite{ruffobook}. 

The $k$TM is defined by the Hamiltonian:
\begin{equation}
H_k = \sum_{i=1}^N\frac{1}{2}\pi_i^2 + V_k (\varphi_1,\ldots,\varphi_N)
\end{equation}
where $\{\varphi_i\}$ are angular variables: $\varphi_i \in [0,2\pi)$,
$\{\pi_i\}$ are the conjugated momenta, and the potential energy $V$ is given by
\begin{equation}
V_k  = \frac{\Delta}{N^{k-1}} \sum_{i_1,...,i_k} [1 - \cos ( \varphi_{i_1}+...+
\varphi_{i_k})] \ ,
\label{ptrigm}
\end{equation}
where $\Delta$ is the coupling constant. 
In what follows only the potential 
energy part will be considered. This interaction energy is apparently of a
mean-field nature, in that each degree of freedom interacts with all the others;
moreover, the interactions are $k$-body ones.

The $k$TM is a generalization of the Trigonometric Model (TM) introduced by 
Madan and Keyes \cite{MK} 
as a simple model for the Potential Energy Surface (PES) -- the hypersurface 
defined by the 
potential energy as a function of the $N$ degrees of freedom -- of simple 
liquids.
The TM is a model for $N$ independent degrees of freedom 
with potential energy
(\ref{ptrigm}) with $k=1$: $V_{k=1}$. 

It shares with Lennard-Jones like systems \cite{lj_sad}
the existence of a regular organization of the critical points of the potential
energy  
above a given minimum (the elevation in energy of the critical points
is proportional to their index) and a regular distribution of
the minima in the configuration space (nearest-neighbor minima lie
at a well defined Euclidean distance).
The PES of the $k$TM maintains the main features of the TM
\cite{ktm},
introducing however a more realistic feature, namely the interaction 
among the degrees of freedom (in the form of a $k$-body interaction).

Using the relation
\begin{equation}
\cos ( \varphi_{i_1}+...+\varphi_{i_k}) = Re ( e^{i\varphi_{i_1}} \cdot \ldots 
\cdot e^{i\varphi_{i_k}} ) \ , 
\end{equation}
the configurational part of the Hamiltonian can be written as
\begin{eqnarray}
V_k &=& N \Delta \left[ 1 - Re (c + i s)^k \right] \nonumber\\
&=&N \Delta \left[ 1 - \sum_{n=0}^{[k/2]}  {k \choose 2n}
\ (-1)^n \ c^{k-2n}\ s^{2n} \right]
\ ,
\label{hamil}
\end{eqnarray}
where $c$ and $s$ are collective variables, functions of $\{\varphi_i\}$:
\begin{eqnarray}
\nonumber
&c& = \frac{1}{N} \sum_{i} \cos  \varphi_i  \ , \\
&s& = \frac{1}{N} \sum_{i} \sin  \varphi_i  \ .
\label{cs}
\end{eqnarray}

We observe also that the model has a symmetry group obtained by the 
transformations
\begin{eqnarray}
\nonumber
&\varphi_i \rightarrow \varphi_i + \ell \frac{2\pi}{k} \\ 
&\varphi_i \rightarrow -\varphi_i 
\end{eqnarray}
If we think of $\varphi_i$ as the angle between a unitary vector in a plane 
and the horizontal axis of this plane, we find that the first transformations
 are rotations in this plane of an angle $\ell \frac{2\pi}{k}$ and the second is 
the reflection with respect to the horizontal axis. This group is also called 
$C_{kv}$.

Let us now derive the thermodynamical properties of the $k$TM.


\subsection{Canonical thermodynamics}

The partition function is 
\begin{equation}
Z_k = \int d\{\varphi\} e^{-\beta H_k} = 
\int d\{\varphi\}  e^{-\beta N \Delta [ 1 - Re (c + i s)^k ] } \ ;
\end{equation}
introducing $\delta$-functions for the variables $c$ and $s$,
\begin{equation}
Z_k = \int d\{\varphi\} \int dx \ dy \ \delta(x-c)\ \delta(y-s)\ 
e^{-\beta N \Delta [ 1 - Re (x + i y)^k ] } \ ,
\end{equation}
and using the integral representation of the $\delta$-function, we obtain for
$Z_k$
\begin{eqnarray}
\nonumber
Z_k = \int d\{\varphi\} \int dx \ dy \ \int N^2 \frac{d\lambda}{2\pi} 
\frac{d\mu}{2\pi}\ 
e^{iN\lambda(x-c)} \  e^{iN\mu(y-s)} \ e^{-\beta \Delta N [ 1 - Re (x + i y)^k ] } = \\
\int dx \ dy \ e^{-\beta \Delta N [ 1 - Re (x + i y)^k ] } 
\int N^2 \frac{d\lambda}{2\pi} \frac{d\mu}{2\pi}\ e^{iN(\lambda x + \mu y)}\ 
\int d\{\varphi\} \ e^{-iN(\lambda c + \mu s)} \ .
\end{eqnarray}
The last integral is easily computable using Eq.\ (\ref{cs}), 
\begin{equation}
\int d\{\varphi\} \ e^{-iN(\lambda c + \mu s)} = \left( \int_0^{2\pi} d\varphi\ 
e^{-i(\lambda \cos  \varphi + \mu \sin  \varphi )} \right)^N \ ,
\end{equation}
and can be written in term of the Bessel function $J_0$   
\begin{equation}
\int_0^{2\pi} d\varphi\ e^{-i\Lambda \cos ( \varphi - \psi )} = 
2\pi\ J_0(\Lambda) \ ,
\end{equation}
where $\lambda=\Lambda\ \cos \psi$, $\mu=\Lambda\ \sin \psi$, and
$\Lambda=\sqrt{\lambda ^2+\mu ^2}$. The partition function can then be written
as
\begin{equation}
Z_k = N^2 (2\pi)^{N-2}
\int dx \ dy \ d\lambda \ d\mu \ 
e^{-N (-i\lambda x - i \mu y +\beta \Delta -\beta \Delta Re (x + i y)^k 
-\log (J_0(\Lambda))} \ ,
\end{equation}
and since the $J_0$ function is always positive, 
there are no problems in defining its logarithm.

We want now to perform a saddle-point evaluation of the integral, so we have
to  look for the minima of the exponent in the complex $\lambda, \mu, x, y$
plane. We note that if that points do not lie on the imaginary axis of the
$\lambda, \mu$ planes, the free energy of the model would be imaginary. So we
can safely rotate the integration path on the imaginary axis in the $\lambda, \mu$
planes, which corresponds to 
the substitutions: $i\lambda \rightarrow \lambda$ and $i\mu \rightarrow \mu$, 
then  $i\Lambda \rightarrow \Lambda$ and $J_0(\Lambda) \rightarrow I_0(\Lambda)$, 
where $I_0$ is 
the modified Bessel-function:
\begin{equation}
I_0(\Lambda) =\frac{1}{2\pi} \int_0^{2\pi} 
d\varphi\ e^{\Lambda \cos  \varphi} \ .
\end{equation}
In conclusion we obtain
\begin{equation}
Z_k = N^2 (2\pi)^{N-2}
\int dx \ dy \ d\lambda \ d\mu \ 
e^{-N g_k(x,y,\lambda,\mu;\beta)} \ ,
\end{equation}
where $g_k$ is the real function
\begin{equation}
g_k(x,y,\lambda,\mu;\beta) = \beta \Delta 
-\lambda x - \mu y  - \beta \Delta \ Re (x + i y)^k 
-\log ( I_0(\Lambda) ) \ .
\label{gp}
\end{equation}
In order to find the stationary points, we first determine the subspace defined
by the equations
\begin{eqnarray}
\frac{\partial g_k}{\partial x} = 0 \ , \\
\frac{\partial g_k}{\partial y} = 0 \ ,
\end{eqnarray}
obtaining the relations
\begin{eqnarray}
\lambda & = & -\beta\ \Delta\ k \ Re (x+iy)^{k-1} \ , \label{eq_lambda}
\\
\mu & = & \beta\ \Delta\ k \ Im (x+iy)^{k-1} \ , 
\label{eq_mu}
\end{eqnarray}
thus we get
\beq
\Lambda = \beta \Delta k |(x+iy)^{k-1}|.
\eeq
Now, using Eqs.\ (\ref{eq_lambda},\ref{eq_mu}), 
we can substitute $\lambda$ and $\mu$ with $x$ and $y$ 
in Eq.\ (\ref{gp}), obtaining, 
in term of the complex number $z=x+iy$,  
\begin{equation}
g_k(z;\beta) = \beta\ \Delta 
+ \beta\ \Delta\ (k-1)\  Re z^k 
-\log  I_0(\beta \Delta p |k^{p-1}|)  \ ,
\end{equation}
and using the polar representation $z= \rho e^{i \psi}$ 
\begin{equation}
g_k(\rho,\psi;\beta) = \beta\ \Delta 
+ \beta\ \Delta\ (k-1)\  \rho^k\ \cos (k\psi) 
-\log  I_0(\beta \Delta k \rho^{k-1})  \ .
\end{equation}
The derivative with respect to $\psi$ leads to
\begin{equation}
-\beta \Delta (k-1) k \rho^k \sin (k\psi) = 0 \ , 
\end{equation}
so that there are $2k$ solutions
\begin{equation}
\psi_n = \frac{n\pi}{k} \ \ (n=1, \ldots, 2k) \ .
\end{equation}
Observing that $\cos (k\psi_n) = (-1)^n$ we obtain
\begin{equation}
g_k(\rho,n;\beta) = \beta\ \Delta 
+ (-1)^n  \beta\ \Delta\ (k-1)\  \rho^k 
-\log  I_0(\beta \Delta k \rho^{k-1})
\end{equation}
and we can restrict ourselves to $n=0,1$. 
Finally, 
the derivative with respect to $\rho$ leads to the stationary points equation:
\begin{equation}
(-1)^n \ \rho = 
\frac{I_1 (\beta \Delta k \rho^{k-1}) }
{I_0(\beta \Delta k \rho^{k-1})} \ ,
\label{stapoint}
\end{equation}
where the modified Bessel-function $I_1$ is defined by
\begin{equation}
I_1(\Lambda) =\frac{1}{2\pi} \int_0^{2\pi} d\varphi \ \cos \varphi \ e^{\Lambda \cos  \varphi} 
= I_0'(\Lambda)\ .
\end{equation}

For $n=1$ we have only the trivial solution $\rho = 0$,  because the $I$
functions are always positive. By using an expansion for small $\rho$ one can
show that this solution is a maximum for $g$.
So we can study only the case $n=0$. 
We note that if there is a non trivial solution (i.e., $\tilde{\rho}(\beta) \neq
0$) of Eq. (\ref{stapoint}), calling $\tilde{g}_k (\beta)$ the value of
$g_k(\beta,\tilde{\rho}(\beta))$, we have
\begin{equation}
Z_k \sim N^2 \ (2\pi)^{N-2}\ e^{-N \tilde{g}_k (\beta) } \ ,
\end{equation}
and the free energy and internal energy are, respectively,
\begin{eqnarray}
f_k(\beta) & = & \beta^{-1} \tilde{g}_k(\beta) - 
\beta^{-1} \log (2\pi)  \ , \\ 
\label{energy}
e_k(\beta) & = & 
\Delta ( 1 - \tilde{\rho}^k) \ .
\end{eqnarray}

Let us now analyze the case $k=1$. In this case
the solutions $\rho=0$ are not present, so that we have only the solution
\begin{equation}
 \tilde{\rho}=\frac{I_1(\beta\Delta)}{I_0(\beta\Delta)} \ . 
\end{equation}
There is no phase transition, and using Eq.\ (\ref{energy}) we have
\begin{equation}
e_1(\beta)=\Delta\left(1-\frac{I_1(\beta\Delta)}{I_0(\beta\Delta)}\right) \ .
\end{equation}
This is the free energy of trigonometric model that has been mentioned 
before.

For $k=2$ the solution $\rho=0$ is stable for high temperatures, 
but a non trivial solution of Eq.\ (\ref{stapoint}) appears at 
$\beta\Delta=1$. The transition temperature is given by the condition 
\beq 
\left. \frac{d^2 g_2(\rho;\beta_c)}{d\rho^2}\right|_{\rho=0}=2\beta_c\Delta(1-\beta_c\Delta)=0 \ , 
\eeq
so that we obtain $\beta_c\Delta=1$; the transition is continuous, 
and the order parameter is $\tilde{\rho}$. 
It is easy to show that $\tilde{x}=\langle c \rangle$ and $\tilde{y}= \langle s
\rangle$  (e.g., by adding an
external field of the form $-N(hc+ks)$ to the Hamiltonian and performing the
limit $h,k \rightarrow 0$); then the vector $(\tilde{x},\tilde{y})$ is the mean
magnetization of the spins represented by the $\varphi_i$. As $\tilde{\rho}$ is
the modulus of the magnetization, for $\beta\Delta>1$, when $\tilde{\rho} \neq
0$, the $C_{2v}$ symmetry is broken. 

When $k> 2$, 
the non
trivial solution of Eq.\ (\ref{stapoint}) appears at a given $\beta'$ but 
becomes stable
only at $\beta''>\beta'$, so that $\tilde{\rho}(\beta)$ and $e(\beta)$
are discontinuous at $\beta''$; instead of the instability region $\beta' <
\beta < \beta''$,
in the microcanonical ensemble  a region where the specific heat
is negative appears, as we shall see below. The $C_{kv}$ symmetry is broken in the
low temperature phase, so that $\tilde{\rho}$ can be used as an order parameter
in revealing the symmetry breaking, even if it is not continuous at $\beta''$.
The transition is then of first order, 
but keeps the symmetry structure of a
second order one, i.e.,  in the low temperature phase 
there are $k$
pure states related by the symmetry group also in the case
of the first order transition. 

%
%

\begin{figure}
\centerline{\psfig{file=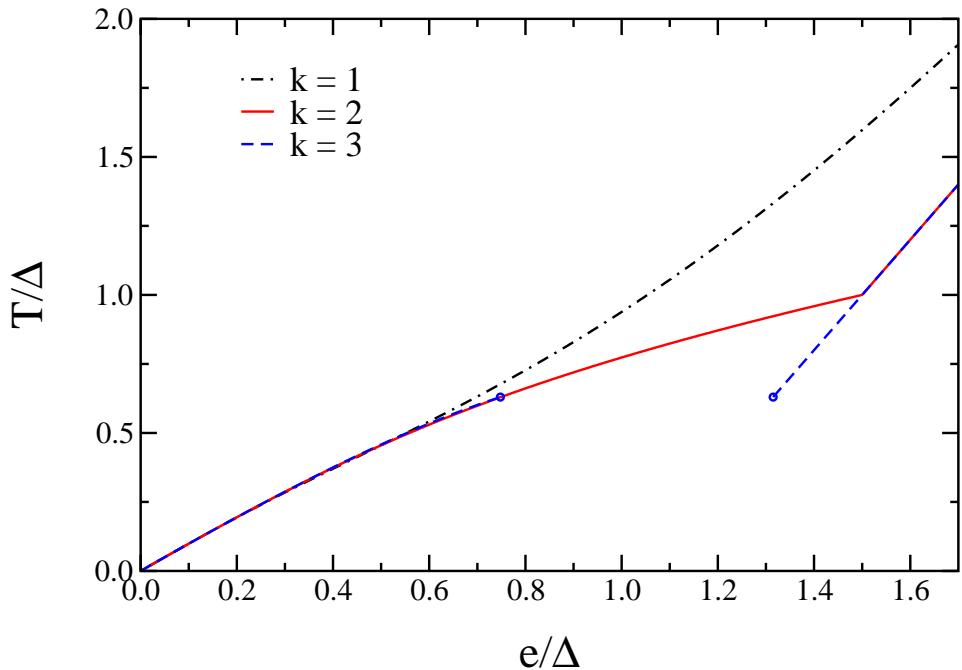,width=14cm,angle=-90}}
\caption{Temperature $T$ as a function of canonical average energy $e$ for three 
different values of $k$; for $k$=1 there is no phase transition, while 
for $k$=2 there is a second order transition and for $k>2$ a first order 
one.}
\label{fig_caloric_can}
\end{figure}

In Fig.~\ref{fig_caloric_can} we report the caloric curve, i.e.,
the temperature $T = \beta^{-1}$ as a function of the average energy (per degree of freedom) $e$,
for three
values of $k$, $k=1$, 2 and 3. As previously discussed, the temperature is an
analytic function of $e$ for $k=1$; for $k=2$ the
system undergoes a second order phase transition at a 
critical temperature $T_c = \Delta$, that changes to first order for $k >2$. 

\begin{figure}
\centerline{\psfig{file=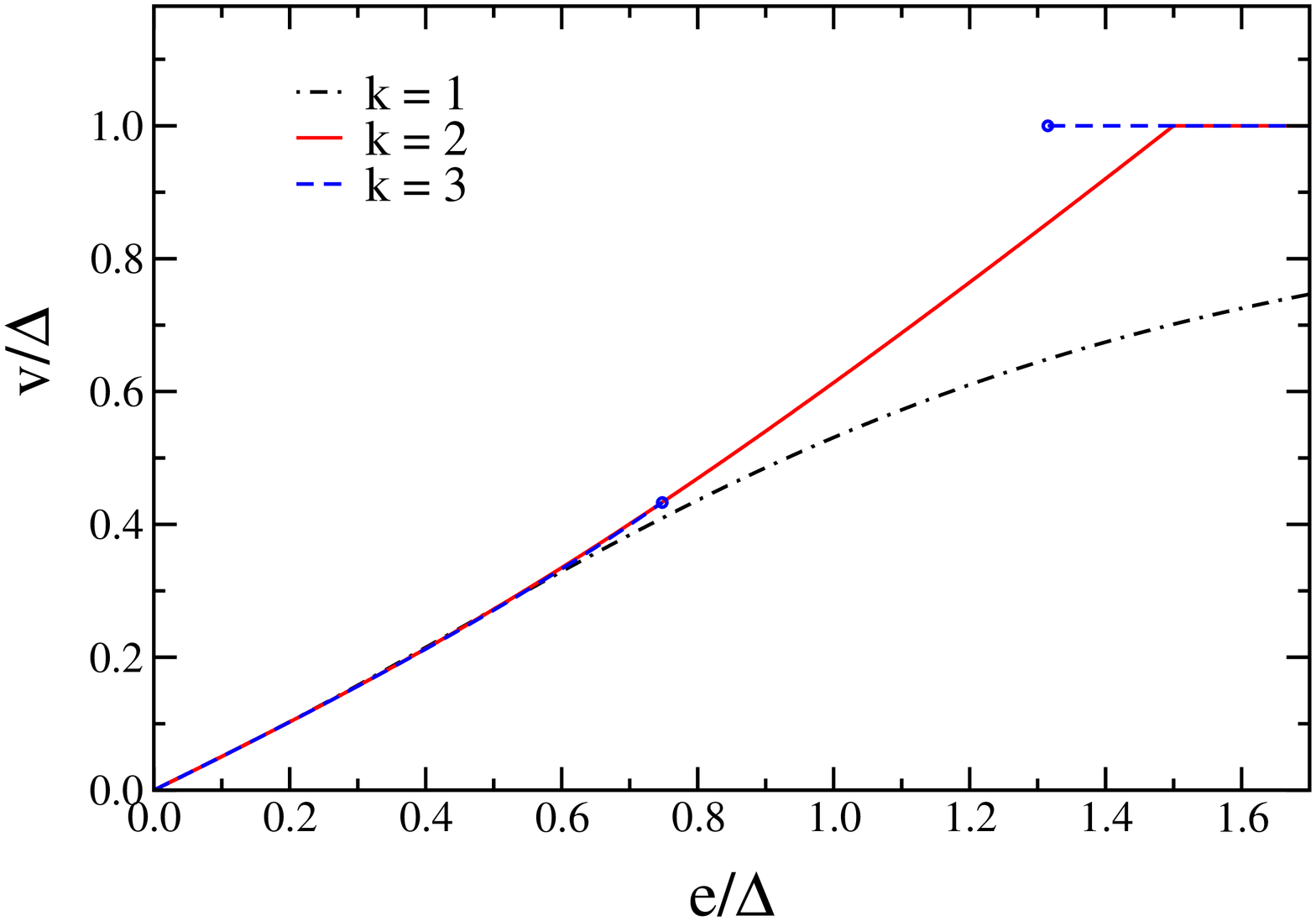,width=14cm,angle=-90}}
\caption{Canonical average potential energy $v$ as a function of canonical
average energy $e$
for $k$=1, 2 and 3. The upper 
phase transition point is, for $\forall k\geq 2$, $v_c=\Delta$.}
\label{fig_ve_can}
\end{figure}

\begin{figure}
\centerline{\psfig{file=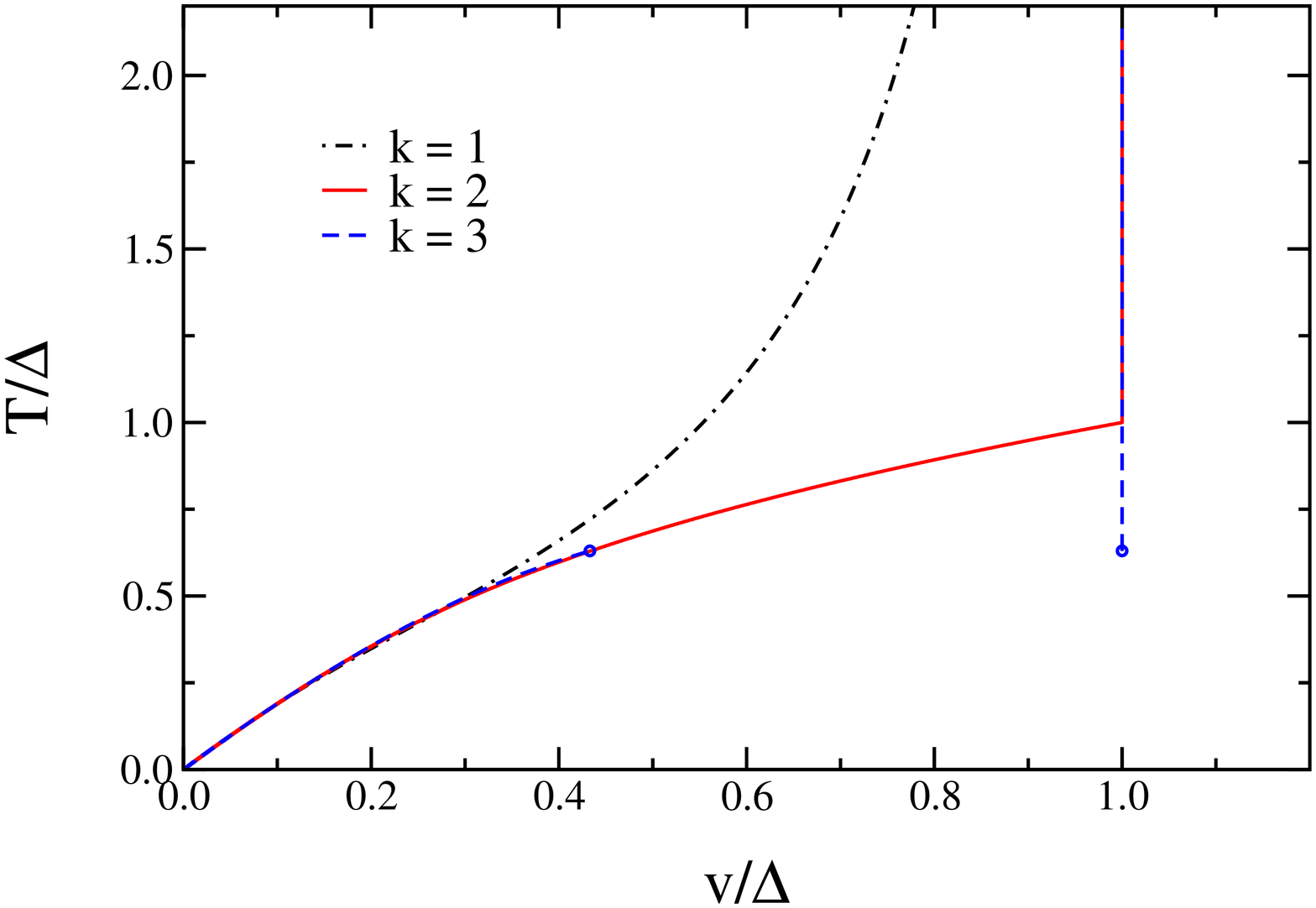,width=14cm,angle=-90}}
\caption{Temperature $T$ as a function of canonical 
average potential energy $v$ for three 
different values of $k$.}
\label{fig_Tv_can}
\end{figure}

In Figs.~\ref{fig_ve_can} and \ref{fig_Tv_can} we report the average
potential energy $v$ as a function of the average energy $e$ and 
the 
temperature $T$ as a function of $v$, respectively. It is apparent that, for
$k\geq 2$, the phase transition point always corresponds to $v_c = \Delta$.

Another feature which shows up in Figs.~\ref{fig_ve_can} and
\ref{fig_Tv_can} is that the average potential energy $v$ never exceeds the
value $\Delta$, i.e., although the maximum of $V/N$ is equal to $2\Delta$, the
region $v > \Delta$ is not thermodynamically accessible to the system. The
reason for this is in the mean-field nature of the system and in the fact that
we are working in the thermodynamic limit $N \to \infty$. According to
Eqs.~(\ref{hamil}), the potential energy can be written as a function of the 
collective variables $c$ and $s$ defined in Eqs.~(\ref{cs}), which are the
components of the function whose statistical average is the 
order parameter, i.e., the ``magnetization''. In the
thermodynamic limit these functions become constants, whose value coincides with
their statistical average, and since $\langle c \rangle = \langle s \rangle = 0$
for $T > T_c$, and from Eqs.~(\ref{hamil}) this implies $v = \Delta$ for all $T
> T_c$.  

As we shall see below, this fact remains true also in the microcanonical
ensemble, which, however, is {\em not} equivalent to the canonical ensemble for
the present model, due to the long-range nature of the interactions.


\subsection{Microcanonical thermodynamics}

As in other simple mean field models, also in the case of the $k$TM it
is possibile to perform a calculation of 
the microcanonical partition function, or microcanonical
density of states in phase space, given
by
\begin{equation}
\Omega_{N,k}(E)=\int \frac{d^N \pi  \ d^N \varphi}{N!} \delta(H_k-E) \ .
\end{equation}
The computation of $\Omega$ is similar to that of $Z$ in the canonical case, so
that we will go through it with less detail.
 
Using the integral representation of the delta function, we get
\begin{equation}
\Omega_{N,k}(E)=\int \frac{d\beta}{2\pi} \int \frac{d^N \pi \ d^N \varphi}{N!} 
e^{-i\beta(H_k-E)} \ .
\end{equation}
Now, as we are looking for a saddle-point evaluation of the integral over 
$\beta$, we can rotate the integration path on the imaginary axis in the 
complex-$\beta$ plane. This is justified because, as in the canonical case, 
the saddle-point is located on this axis.
We can now perform the integration over the momenta and use the fact  that 
$V_k(\varphi)=V_k(c(\varphi),s(\varphi))$, see Eq.\ (\ref{hamil}), to obtain
\begin{eqnarray}
\nonumber
\Omega_{N,k}(E)=&{\cal C}_N \ \rho^N \int d\beta \ d\xi \ d\eta \ 
\beta^{-\frac{N}{2}} \ e^{\beta(E-V_k(\xi,\eta))} \cdot \\ &\cdot \int d^N
\varphi \ \delta[N(\xi-c(\varphi))] \ \delta[N(\eta-s(\varphi))] \ ,
\end{eqnarray}
where $\rho =N/L$ and the constant ${\cal C}_N$ gives only a constant 
contribution to the entropy per particle, i.\ e., it is at most of order $e^N$.
The last integral can be evaluated using again the integral representation of 
the delta function, and rotating then the  integration path as previously 
discussed; it turns out to be:
\[
\int \frac{d\mu \ d\nu}{(2\pi)^2} e^{-N(\mu \xi + \nu \eta)} \int d^N\varphi 
\ e^{\sum_i (\mu \cos \varphi_i + \nu \sin\varphi_i)} = 
\]
\[
= \int \frac{d\mu \ d\nu}{(2\pi)^2} e^{-N(\mu \xi + \nu \eta)} 
(2\pi I_0(\Lambda))^N \ ,
\]
having defined $\Lambda =\sqrt{\mu^2+\nu^2}$; $I_0$ is a Bessel function as
before.
We can then write the density of states as
\begin{equation}
\Omega_{N,k}(e)= {\cal C}_N \ \rho^N \int d {\mathbf{m}} \ 
e^{N f_k({\mathbf{m}},e)} \ ,
\end{equation}
where $\mathbf{m}\equiv (\beta,\xi,\eta,\mu,\nu)$, $e =E/N$ and
\[
f_k({\mathbf{m}},e)=\beta e-\beta \Delta [1-\text{Re}(\xi+i\eta)^k] 
\]
\[ - 
\frac{1}{2} \log \beta - \mu \xi -\nu \eta + \log I_0(\Lambda) \ .
\]
Then, using the saddle-point theorem, the entropy per particle, $s =S/N$, 
is given by ($k_B$=1):
\begin{equation}
s_k(e)=\lim_{N \rightarrow \infty} \frac{1}{N} \log
\Omega_{N,k}(e)=\max_{\mathbf{m}} f_k({\mathbf{m}},e) \ .
\end{equation}
To find the maximum of $f_k(\mathbf{m},e)$ one can calculate analytically
some derivatives of $f$ to obtain a one-dimensional problem that can be easily
 solved numerically with standard methods.

\begin{figure}
\centerline{\psfig{file=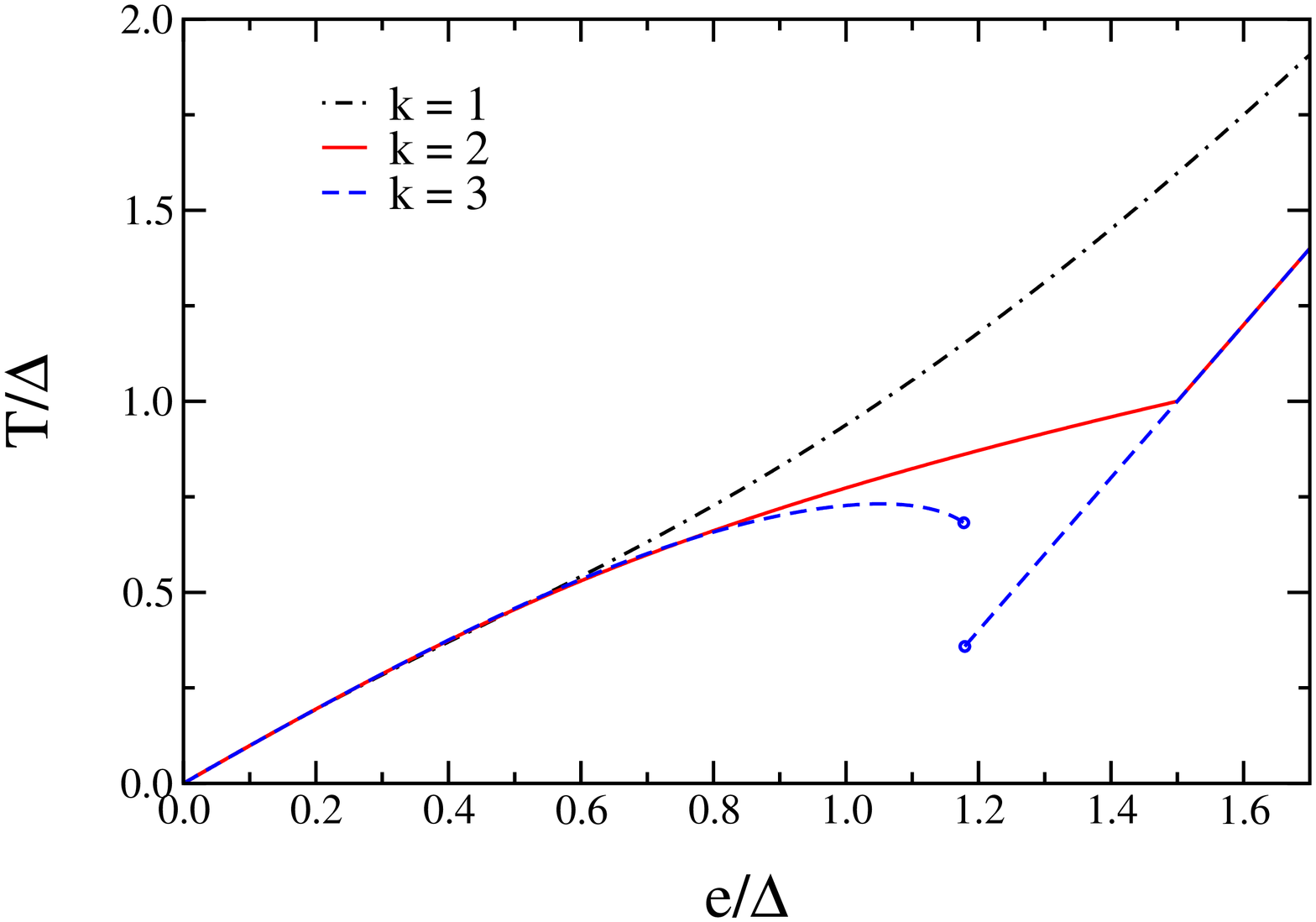,width=14cm,angle=-90}}
\caption{Microcanonical temperature $T$ as a function of energy $e$ for three 
different values of $k$; for $k$=1 there is no phase transition, while 
for $k$=2 there is a second order transition and for $k>2$ a first order 
one.}
\label{fig_caloric_micro}
\end{figure}

As already done in the case of the canonical ensemble, 
in Fig.~\ref{fig_caloric_micro} we report the microcanonical 
caloric curve, i.e.,
the temperature $T$ as a function of the energy (per degree of freedom) $e$,
$T(e)=[{\partial s}/{\partial e}]^{-1}$ for three
values of $k$, $k=1$, 2 and 3. As in the canonical case, 
the temperature is an
analytic function of $e$ for $k=1$, while for $k$=2 the
system undergoes a second order phase transition at a certain
energy value $e_c$, that changes to first order for $k >2$. 

We note that, for $k >
2$, in a region of energies smaller than the critical energy $e_c$ of the
first-order phase transition 
the curve $T(e)$ has a negative slope, i.e., the system has a negative
specific heat. The $k$TM is then another physical model where this feature is
found (see, e.g., Ref.\ \cite{ruffobook} for other examples). 
This is not surprising at all since we are considering the
\emph{microcanical} thermodynamics of a system with long-range interactions;
such a region is {\em not} present when we consider the canonical
ensemble, as shown above; there, the region of negative specific heat
corresponds to the region of instability of the non-trivial solution of the
saddle-point equations.  

\begin{figure}
\centerline{\psfig{file=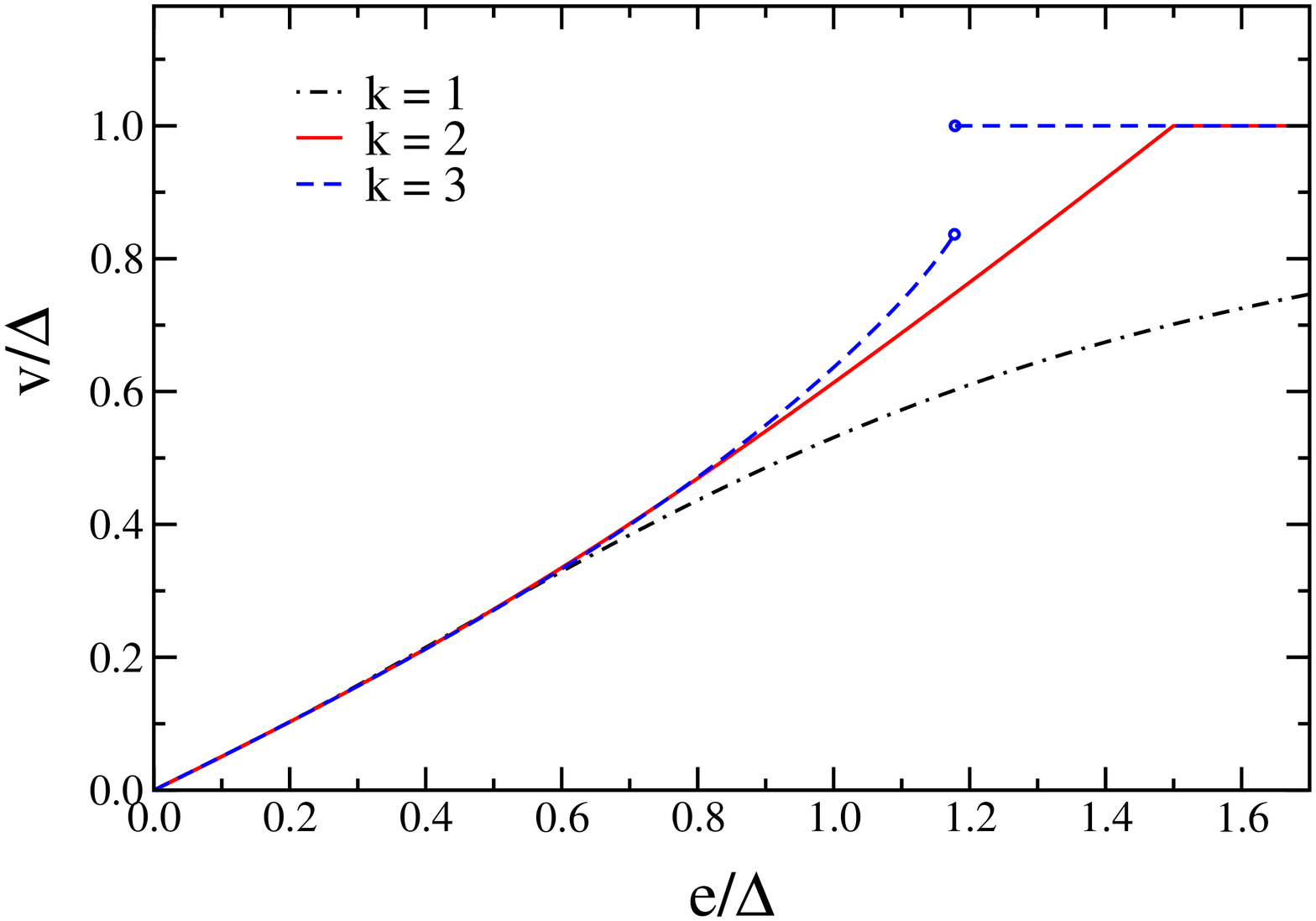,width=14cm,angle=-90}}
\caption{Microcanonical average potential energy $v$ as a function of energy $e$
for $k$=1, 2 and 3. The 
phase transition point is, for $\forall k\geq 2$, $v_c=\Delta$.}
\label{fig_ve_micro}
\end{figure}

\begin{figure}
\centerline{\psfig{file=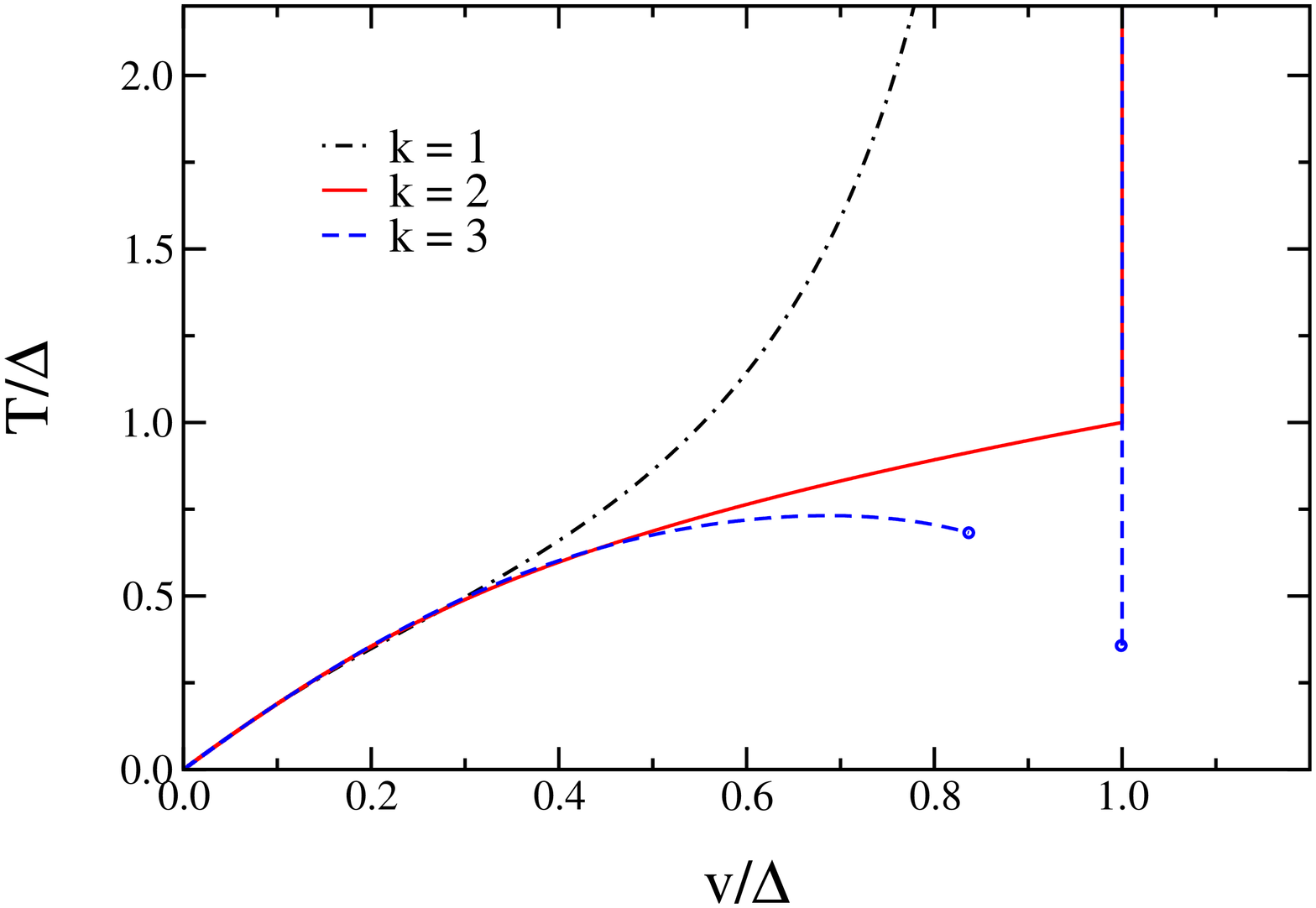,width=14cm,angle=-90}}
\caption{Microcanonical temperature $T$ as a function of microcanonical 
average potential energy energy $v$ for three 
different values of $k$.}
\label{fig_Tv_micro}
\end{figure}

In Figs.~\ref{fig_ve_micro} and \ref{fig_Tv_micro} we report the average
microcanonical potential energy $v$ as a function of $e$ and the microcanonical
temperature $T$ as a function of $v$, respectively. It is apparent that, for
$k\geq 2$, the phase transition point always corresponds to $v_c = \Delta$.

As in the canonical case, the average potential energy $v$ never exceeds the
value $\Delta$, i.e., the
region $v > \Delta$ is not thermodynamically accessible to the system also in
the microcanonical ensemble. 

\subsection{Topology of configuration space}

In this Section we want to investigate the relation between the phase
transitions occurring in the $k$TM and the topology of its configuration space.
Given the potential energy $V$, the following submanifolds of 
configuration space are defined:
\[
M_{v}\equiv\{\varphi \ | \ V(\varphi) \leq Nv\} \, .
\]
As $v$ varies from the minimum to the maximum allowed value of $V(\varphi)/N$, 
the
manifolds $M_v$ progressively cover the whole configuration space. These
submanifolds, or their boundaries $\Sigma_v = \partial M_v$, are then a possible
way to depict the potential energy landscape of the system.  The topology
of the $M_v$'s can be studied using Morse theory \cite{morse}: whenever a
\emph{critical value} of $V(\varphi)/N$ -- i.e., a value corresponding to one or
more \emph{critical points}, where the differential of $V(\varphi)/N$ vanishes
-- is crossed, the topology of the $M_v$'s change. It has been conjectured
\cite{cccp,physrep} that some of these topology changes are the ``deep'' cause
of  the
presence of a phase transition. The correspondence between major topology
changes of the $M_v$'s and $\Sigma_v$'s and phase transitions has been checked
in two particular models \cite{phi4,xy}; more recently, it has been proved
\cite{theorem} that
a topology change is a \emph{necessary} condition for a phase transition under
rather general assumptions\footnote{Mean-field models with long-range
interactions, like those to be analyzed in the present paper, 
do not rigorously fulfill the hypotheses of the theorem proved in
\protect\cite{theorem}; the fact that we do find a correspondence between the
phase transition and a particular topology change also for these models, as
already found in other models with long-range interactions, suggests that the
theorem could be probably extended to a wider class of systems, which however
would not necessarily include all the possible kinds of models with 
long-range, mean-field-like interactions (see also the discussion in
Sec.~\protect\ref{conclusions}).}, but the sufficiency
conditions are still lacking.

A natural way to characterize topology changes involves the computation of some
\emph{topological invariants} of the manifolds under investigation. One of such
invariants is the Euler characteristic $\chi$: 
in \cite{xy,phi4} it was 
shown that the Euler characteristic $\chi(v)$ of the submanifolds 
$M_v$ and/or of the $\Sigma_v$ shows a 
singularity in correspondence of the potential energy value $v_c=v(e_c)$ at 
which the transition takes place: this seems then to signal a particularly
``strong'' topology change. 
Remarkably, the Euler characteristic of $M_v$ can be calculated 
{\it analytically}
in our model. The general definition is \cite{nakahara}:
\begin{equation}
\chi(v) \equiv \chi(M_v)=\sum_{i=0}^{N} (-1)^i \mu_i(M_v) \ ,
\label{defchi}
\end{equation}
where the \emph{Morse numbers} $\mu_i(M_v)$ 
are the number of critical 
points of index $i$ of the function $V(\varphi)$ 
that belong to the manifold $M_v$.
As already mentioned, the critical points (called \emph{saddles} in other
contexts, e.g., in the physics of glasses) $\tilde{\varphi}$ are defined by the
condition $dV_k(\tilde{\varphi})=0$, and their index $i$ (otherwise called the
\emph{order} of the saddle) is defined as
the number of negative eigenvalues of the Hessian matrix
\beq
{\cal H}^k_{ij}(\tilde{\varphi})=\left. \left(\frac{\partial^2 V_k}{\partial \varphi_i
\partial \varphi_j}\right) \right|_{\tilde{\varphi}}\, .
\eeq 
To be valid, Eq.~(\ref{defchi}) requires that $V(\varphi)$ is a Morse function,
i.e., that its critical points are nondegenerate: this means that all the
eigenvalues of the Hessian are nonzero at a critical point and that the critical
points themselves are isolated.
 
To determine the critical points we have then to solve the system
\beq
\frac{\partial V_k}{\partial \varphi_j}= 0 ~~~~~~ \forall\, j=1,\ldots,N
\eeq
that is, inserting Eq.~(\ref{hamil}) in the equations above,
\beq
-\Delta \ k \ \text{Re}[i(c+is)^{k-1} 
e^{i \varphi_j}] = 
\Delta \ k \ \zeta^{k-1} \sin[(k-1)\psi + \varphi_j] = 0 ~~~ \forall\,
j=1,\ldots,N \, ,
\label{saddef}
\eeq
where we defined $c+is=\zeta e^{i \psi}$. From Eq.~(\ref{hamil}) we have
\beq
V_k(\varphi)=N\Delta[1-\zeta^k \cos(k\psi)]~;
\eeq
then all the critical points with 
$\zeta(\tilde{\varphi})=0$ have energy $v=V(\tilde{\varphi})/N=\Delta$. 
We note that they
correspond to vanishing magnetization.
Let us now consider all the critical points with $\zeta(\tilde{\varphi})\not=0$. 
Then Eq.~(\ref{saddef}) becomes
\begin{equation}
\sin [(k-1)\psi + \varphi_j]=0 ~~~~~~ \forall\, j=1,\ldots,N \, ,
\label{sadpartial}
\end{equation}
and its solutions are
\begin{equation}
\label{sadcond}
\tilde{\varphi}_j^{\mathbf{m}}=[m_j \pi - (k-1) \psi]_{\mathrm{mod} \ 2 \pi} ~ ,
\end{equation}
where $m_j\in\{ 0,1 \}$. Since in Eq.~(\ref{saddef}) $\zeta$ appears to the
$k-1$-th power, in the case $k=1$ Eqs.~(\ref{saddef}) and (\ref{sadpartial})
coincide. This means that the solutions given in Eq.~(\ref{sadcond}) are
\emph{all} the critical points, regardless of their energy, in the case $k=1$
and all the critical points \emph{but} those with energy $v=\Delta$ in the case
$k > 1$.    
The critical point $\tilde{\varphi}^{\mathbf{m}}$ is then characterized 
by the set 
$\mathbf{m}\equiv\{ m_j \}$.
To determine the unknown constant $\psi$ we have to substitute 
Eq. (\ref{sadcond}) in the self-consistency equation
\begin{equation} \label{selfcons}
\zeta e^{i\psi}=c+i s=N^{-1} 
\sum_j
e^{i\varphi_j}=N^{-1} e^{-i\psi(k-1)} 
\sum_j
 (-1)^{n_j} \, .
\end{equation}
If we introduce the quantity $n(\tilde{\varphi})$ defined by
\begin{equation}
\label{frac_ord}
n = N^{-1}
\sum_j m_j ~,
\end{equation}
which means 
\beq
1-2n=N^{-1} 
\sum_j
 (-1)^{n_j} \, ,
\eeq
we have from Eq. (\ref{selfcons})
\begin{eqnarray}
&\label{z-x}
\zeta=|1-2n| \ , \\
&\psi_l=
\begin{cases}
2l\pi /k \hspace{2cm} \text{ for } n < 1/2 \ , \\
(2l+1)\pi /k \hspace{1.12cm} \text{ for } n > 1/2 \ , 
\end{cases}
\label{rhoepsi}
\end{eqnarray}
where $l \in\Bbb{Z}$. 
Then the choice of the set $\{m_j\}$ is not sufficient to specify the set 
$\{\varphi_j\}$, because the constant $\psi$ can assume some different values. 
This fact is connected with the symmetry structure of the potential energy 
surface: the different values of $\psi_l$ correspond to the symmetry-related
critical points under the group $C_{kv}$.

We can then state that all the critical points with $\zeta \neq 0$ -- whose  
energy $v\neq\Delta$ -- have the 
form
\begin{equation}
\label{sadfin}
\tilde{\varphi}_j^{\mathbf{m},l}=[m_j \pi - (k-1) \psi_l]_{\text{mod} \ 2 \pi} \ .
\end{equation}
The Hessian matrix is given by
\begin{eqnarray} \nonumber
{\cal H}^k_{ij}=&
\Delta\ k\ Re [N^{-1} (k-1) (c+is)^{k-2} e^{i(\varphi_i+\varphi_j)}  \\
&+\delta_{ij} (c+is)^{k-1} e^{i \varphi_i)} ]  \ .
\end{eqnarray}
In the thermodynamic limit it becomes diagonal 
\begin{equation}
\label{hessdiag}
{\cal H}^k_{ij}=
\delta_{ij}\ \Delta\ k\ \zeta^{k-1}\
\cos \left( \psi(k-1)+\varphi_i \right) \ .
\end{equation}
One can not {\it a priori} neglect the contribution of the off-diagonal terms 
to the eigenvalues of ${\cal H}$, but we have numerically checked that their 
contribution change at most the sign of only one eigenvalue over $N$: we note
that in the case of the mean-field $XY$ model this result has been proven
explicitly \cite{xy}.
Neglecting the off-diagonal contributions, the eigenvalues of the Hessian 
calculated in the critical point $\tilde{\varphi}$ are obtained substituting 
Eq.~(\ref{sadfin}) in Eq.~(\ref{hessdiag}):
\begin{equation}
\lambda_j = (-1)^{m_j} \Delta\ k\ \zeta^{k-1} \ ,
\end{equation}
so the index of the critical point is simply the number of $m_j=1$ in 
the set $\mathbf{m}$; we 
can identify the quantity $n(\tilde{\varphi})$ given by Eq.~(\ref{frac_ord}) 
with the \emph{fractional index} $\nu/N$ of the critical point $\tilde{\varphi}$.
Then, from Eq.~(\ref{hamil}), (\ref{z-x}) and (\ref{rhoepsi}) we get a 
relation between the fractional index $n(\tilde{\varphi})$ and the potential 
energy $v(\tilde{\varphi})=V(\tilde{\varphi})/N$ at each critical point 
$\tilde{\varphi}$:
\begin{equation}
\label{x-v}
n(v)=\frac{1}{2}
\left[1-\text{sgn}\left(1-\frac{v}{\Delta} \right)
\left|1-\frac{v}{\Delta}\right|^{1/k} \right] \ ,
\end{equation}
Moreover, the number of critical points of given  
index $\nu$ is simply the number of 
way in which one can choose $\nu$ times 1 among the $\{m_j \}$, see 
Eq.~(\ref{sadfin}), multiplied for a constant ${\cal A}_k$ that takes into 
account the degeneracy introduced by Eq.~(\ref{rhoepsi}).

We have then completely characterized the critical points with $\zeta \neq 0$.
Now we are going to argue that, in order to compute the Euler characteristic of
the manifolds $M_v$, we can neglect the critical points with $\zeta =
0$, thus showing that the knowledge of the critical points considered so far is
sufficient. The critical points with $\zeta = 0$ 
are degenerate: the Hessian vanishes at these points. This
means that the potential energy is no longer a proper Morse function when $v
\geq \Delta$,
and we could use its critical points to compute the Euler characteristic of the
manifolds $M_v$ only when $v < \Delta$. To overcome this difficulty we reason as
follows. Morse functions are dense in the space of smooth functions: this means,
in practice, that if a function $f$ is {\em not} a Morse function, any small
perturbation will make it a proper Morse function \cite{MorseCairns}, and we can
consider, as our Morse function, the function $\tilde{V}_k$, i.e., 
the potential energy plus a linear term which
can made as small as we want:  
\beq
\tilde{V}_k = V_k + \sum_{i=1}^N h_i \varphi_i~,
\eeq
where $h \in {\Bbb R}^N$. The perturbation changes only slightly the critical
points with $\zeta \neq 0$, but completely removes the points with $\zeta = 0$
for any $h \not=0$, no matter how small. All the critical points 
of this function are given by the
solutions of the equations 
\begin{equation}
\sin [(k-1)\psi + \varphi_j]=h_j ~~~~~~ \forall\, j=1,\ldots,N \, ,
\label{sadpartial_pert}
\end{equation}
which are only a slight deformation of Eqs.\ (\ref{sadpartial}), so that
provided all the $h$'s are very small
the numerical values of critical points and critical levels will essentially 
coincide with
those computed so far, in the case $h = 0$ but assuming $\zeta \not = 0$.

The fractional index
$n=\nu/N$ of the critical points is a well defined
monotonic function of their potential energy $v$, 
given by Eq. (\ref{x-v}), and 
the number of critical points of a given index $\nu$ is ${\cal A}_k {N \choose
\nu}$.
Then the Morse indexes $\mu_\nu(M_v)$ of the manifold
$M_v$ are given by ${\cal A}_k {N \choose \nu}$ if $\nu/N\leq n(v)$ and 0
otherwise, and the Euler characteristic is
\begin{equation}
\label{chiv}
\chi(v) ={\cal A}_k \sum_{\nu=0}^{Nn(v)} (-1)^\nu  \ {N
\choose \nu} ={\cal A}_k (-1)^{N n(v)} {N-1 \choose N n(v)} \ ,
\end{equation}
using the relation 
$\sum_{\nu=0}^m (-1)^N {N \choose \nu}=(-1)^m {N-1 \choose m}$. 

\begin{figure}
\centerline{\psfig{file=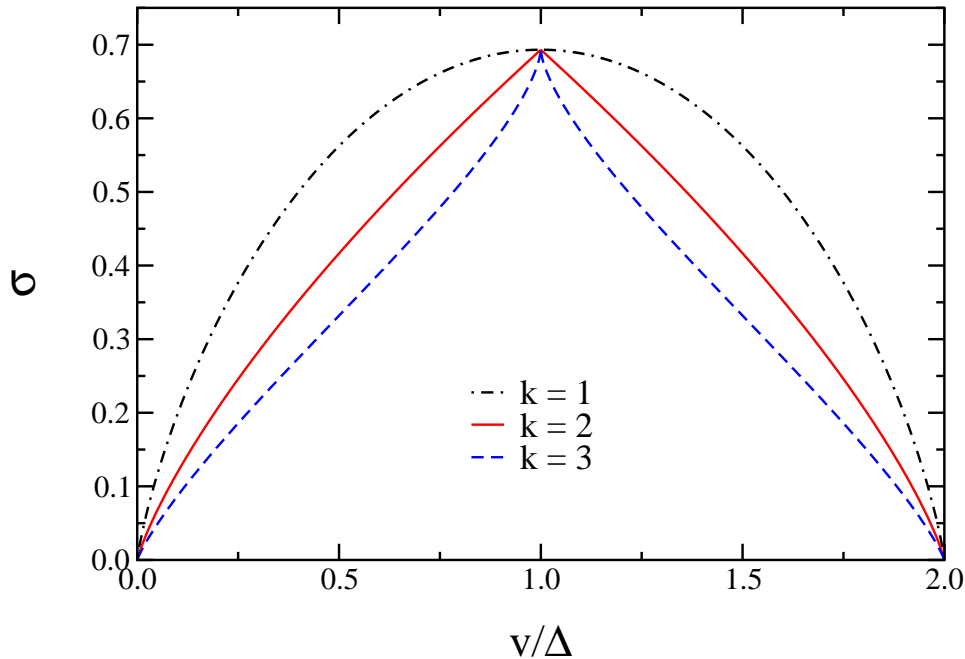,width=14cm,angle=-90}}
\caption{Logarithmic Euler characteristic of the $M_v$ manifolds $\sigma(v)$ 
(see text) as a function of the potential energy $v$. The 
phase transition is signaled as a singularity of the first derivative at 
$v_c =\Delta$; the sign of the second derivative around the singular point 
allows to predict the order of the transition. The region $v>\Delta$, in 
which $\sigma'(v)<0$, in not reached by the system (see text).}
\label{fig_sigma}
\end{figure}

In Fig.\ \ref{fig_sigma} we plot $\sigma(v)=\lim_{N\rightarrow \infty}
\frac{1}{N}\log | \chi(v) |$, that, from Eq.\ (\ref{chiv}), is given by:
\begin{equation}
\sigma(v)=  -n(v) \log n(v) -(1-n(v)) \log (1-n(v)) \ .
\end{equation}
It has to be
stressed that $\sigma(v)$ is a purely {\it topological} quantity,
being related only to the properties  of the potential energy
surface defined by $V_k(\varphi)$, and, in particular, to the energy
 distribution of its saddle points.
From Fig.~\ref{fig_sigma} we can see that there is an evident  
signature of the phase transition in 
the analytic properties of $\sigma(v)$. First, we observe that the region 
$v>\Delta$ is never reached by the system, as discussed before and 
showed in Figs. \ref{fig_ve_micro}
and \ref{fig_Tv_micro} as to the microcanonical case, and in 
Figs.~\ref{fig_ve_can} and \ref{fig_Tv_can} as to the canonical case; 
this region is characterized by $\sigma'(v)<0$. 
The main observations are that: $i)$ for
$k$=1, where there is no phase transition, the function $\sigma(v)$
is analytic;
$ii)$ for $k$=2, when we observe a second order phase transition, 
the first derivative of
$\sigma(v)$ is discontinuous at $v_c=v(e_c)=\Delta$, and its second derivative 
is {\it negative} around the singular point. $iii)$ for $k\geq 3$ the first 
derivative of $\sigma(v)$ is also discontinuous at the transition point 
$v_c=\Delta$, but its second derivative is {\it positive} around $v_c$. 
In this case {\it a first order transition takes place}.
Therefore the investigation of the potential energy topology, via $\sigma(v)$,
 allows us to establish not only the location but also the order of the phase
 transitions, without 
introducing any statistical measure.

The previous results suggest us to conjecture that there is a relation between 
the thermodynamic entropy of the system and the topological properties of the
potential energy landscape, as probed by $\sigma(v)$. We recall that 
the presence
 of a first order transition with a discontinuity in the energy is generally 
related \cite{gallavotti} to a region of negative specific heat, i.e., of 
positive second derivative of the entropy. Thus, it seems that 
at least around the transition point the thermodynamic entropy and 
$\sigma(v(e))$ are closely related: more precisely, it seems that  
the jump in the second 
derivative of $s(e)$ is determined by the jump in the second derivative of 
$\sigma(v(e))$. Then it should be possible to write
\begin{equation}
s(e) \sim \sigma(v(e)) + {\cal R}(e)
\end{equation}
where ${\cal R}(e)$ is analytic (or, at least, $C^2$) around the transition 
point.

In Sec.\ \ref{directlink} we explain how such a link between thermodynamics and
topology could be obtained. But before doing so we show a different way of
looking at the topology of the submanifolds of configuration space defined by
the potential energy.

\subsection{Topology of the order parameter space}

A feature of many mean-field models (although not of all of them) is that the
potential energy can be written as a function of a collective variable,
whose statistical average is the order parameter. In the case of the
$k$-trigonometric model this variable is the two-dimensional ``magnetization''
vector defined as $\mathbf{m} = (c,s)$, where (see Eqs.\ \ref{cs}) 
\beq
\left \{ \begin{array}{ccc} 
c & = & {\displaystyle \frac{1}{N}\sum_{i=1}^N \cos\left(\varphi_i \right)\,} ,\\
& & \\
s & = & {\displaystyle \frac{1}{N}\sum_{i=1}^N \sin\left(\varphi_i \right)\,} . 
\end{array}
\right.
\eeq 
Written in terms of $(c,s)$, the potential energy is a function defined on the
unitary disk in the real
plane, which is given by (see Eq.\ (\ref{hamil}))
\beq
V_k (c,s) = N \Delta \left[ 1 - \sum_{n=0}^{[k/2]}  {k \choose 2n}
\ (-1)^n \ c^{k-2n}\ s^{2n} \right]~.
\label{V(cs)}
\eeq
In the particular cases $k = 1,2,3$ the potential energy $V_k$ reads as
\begin{eqnarray}
V_1(c,s) &=& N \Delta (1 - c)~ , \\
V_2(c,s) &=& N \Delta (1 - c^2 + s^2)~ , \\
V_3(c,s) &=& N \Delta (1 - c^3 + 3cs^2)~ ,
\end{eqnarray}
and it is then natural to investigate the topology of the $M_v$'s seen as
submanifolds of the unitary disk in the plane, i.e., we now consider the
submanifolds
\beq 
{\cal M}_v \equiv \{ (c,s) \in D^2 \ | \ V_k(c,s) \leq Nv\} \, .
\eeq
where $D^2 \equiv \{ (c,s) \in \mathbb{R}^2 \ | \ c^2 + s^2 \leq 1\}$.
The ${\cal M}_v$'s are nothing but the $M_v$'s projected onto the
``magnetization'' plane.

\begin{figure}
\centerline{\psfig{file=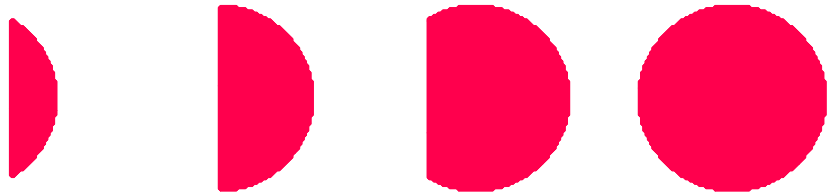,width=14cm}}
\caption{The submanifolds ${\cal M}_v$ in the case $k = 1$ for $v =
0.5\Delta,\Delta,1.5 \Delta,2 \Delta$
(from left to right). All the submanifolds are topologically equivalent to a
single disk.}
\label{fig_mag_1}
\end{figure}

\begin{figure}
\centerline{\psfig{file=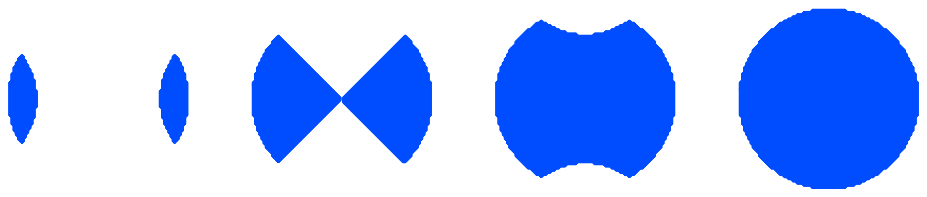,width=14cm}}
\caption{The submanifolds ${\cal M}_v$ in the case $k = 2$ for $v =
0.5\Delta,\Delta,1.5 \Delta,2 \Delta$
(from left to right). As $v < v_c = \Delta$ the submanifolds are topologically
equivalent to two disconnected disks, while as $v > v_c$ they are equivalent to a
single disk.}
\label{fig_mag_2}
\end{figure}

\begin{figure}
\centerline{\psfig{file=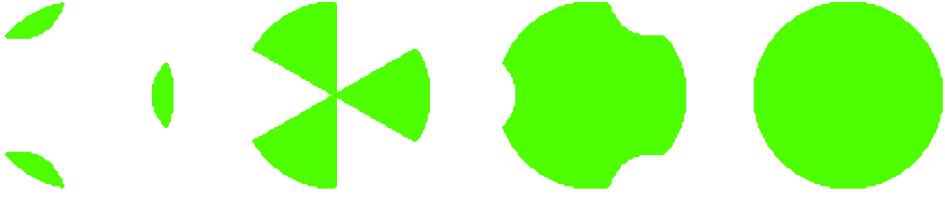,width=14cm}}
\caption{The submanifolds ${\cal M}_v$ in the case $k = 3$ for $v =
0.5\Delta,\Delta,1.5 \Delta,2 \Delta$
(from left to right). As $v < v_c = \Delta$ the submanifolds are topologically
equivalent to three disconnected disks, while as $v > v_c$ they are equivalent to a
single disk.}
\label{fig_mag_3}
\end{figure}

The topology of these manifolds can be studied
directly, by
simply drawing them. In the case $k = 1$, where no phase transition is present,
no topology changes occur in the ${\cal M}_v$'s, i.e., all of them are
topologically equivalent to a single
disk $D^2$ (Fig.\ \ref{fig_mag_1}). When $k = 2,3$, and a phase transition is
present, there is a topology change precisely at $v_c = \Delta$, where $k$ disks
merge into a single disk (see Figs.\ \ref{fig_mag_2} and \ref{fig_mag_3}). The
detail of the transition, i.e., the number of disks which merge into one,
clearly reflects the nature of the symmetry breaking for the particular value of
$k$ considered (similar pictures are obtained for $k > 3$). 
Thus, when projected onto the order parameter space, the
correspondence between topology changes and phase transitions becomes one-to-one
(this was already found in \cite{xymf} for the mean-field $XY$ model); however, at
variance with the study of the topology of the ``full'' $M_v$'s, no
direct way to discriminate between first- and second-order transitions seems
available in this picture.  

\section{Topology and thermodynamics: a direct link}

\label{directlink}

We now show how it is possible to establish
a general relationship between topology and
thermodynamics. This can be 
achieved by improving some preliminary results on the
subject reported in \cite{franzotesi,CSCP}.

Consider a generic classical system described by a  
Hamiltonian 
\[
H = \frac{1}{2}\sum_{i=1}^N p_i^2 + V(q)
\]
where $q=(q_1,\dots,q_N)$ and the symbols have standard meaning. 
Then consider the microcanonical entropy  $S(E)$ defined as
\begin{equation}
S(E)=\frac{k_B}{N} \log \Omega_\nu (E)~,
\label{Entropia}
\end{equation}
where $\nu =2N-1$, where we shall put $k_B=1$ and where 
\begin{equation}
\Omega_{\nu}(E)=\frac{1}{N!}\int_{\Sigma_E}\frac{d\sigma}{\Vert\nabla H\Vert},
\label{volmu}
\end{equation}
with 
$\Vert\nabla H\Vert =\{\sum_i p_i^2 + [\nabla_iV(q)]^2\}^{1/2}$, i.e.,
$\Omega_\nu$ is the microcanonical density of states. Here
$\Sigma_E$ is the constant-energy hypersurface in the $2N$-dimensional
phase space $\Gamma$ corresponding to the total energy $E$, that is 
$\Sigma_E=\{(p_1,\dots,p_N,q_1,\dots,q_N)\in\Gamma \vert H(p,q)=E\}$.

From the general derivation formula \cite{laurence}
\begin{equation}
{d^k \over dE^k}  \int_{\Sigma_E} \alpha~d\sigma 
        = \int_{\Sigma_E} A^k(\alpha)\, d\sigma~~,
\end{equation}
where $\alpha$ is an integrable function and $A$ is the operator
\[
        A(\alpha)= {\nabla \over  \Vert \nabla H\Vert}
        \cdot \left( \alpha \cdot
       {\nabla H \over \Vert \nabla H\Vert} \right)
\]
the following result is worked out 
\begin{equation}
\frac{d\Omega_{\nu}(E)}{dE} = 
 \frac{1}{N!}
\int_{\Sigma_E} \frac{d\sigma}
{\Vert\nabla H\Vert}\, \frac{M_1^\star}{\Vert\nabla H\Vert}+
{\cal O}\left(\frac{1}{N}\right) ~~ ,
\label{ballerotte}
\end{equation}
where $M_1^\star = \nabla (\nabla H/ \Vert\nabla H\Vert)$. $M_1^\star$ is 
directly
proportional to the mean curvature $M_1$ of $\Sigma_E$ seen as a
submanifold of $\mathbb{R}^N$ \cite{docarmo} 
according to the simple relation $M_1=-M_1^\star/(2N-1)$. 
By integrating equation Eq.~(\ref{ballerotte})
we obtain the following equivalent expression for $\Omega_{\nu}(E)$
\begin{eqnarray}
\Omega_\nu(E)=\frac{1}{N!}\int_0^E d\eta  \int_{\Sigma_\eta}\frac{d\sigma}
{\Vert\nabla H\Vert}\, \frac{M_1^\star}{\Vert\nabla H\Vert}
=\frac{1}{N!}\int_{M_E}d \mu \ \frac{M_1^\star}{\Vert\nabla H\Vert}
\label{coarea1}
\end{eqnarray}
and then, at large $N$, 
considering that the volume measure $d \mu$ concentrates on the
boundary $\Sigma_E$, we write
\begin{eqnarray}
\frac{1}{N!}\int_{M_E}d \mu \ \frac{M_1^\star}{\Vert\nabla H\Vert}& \simeq &
\frac{(\delta E)}{N!}\int_{\Sigma_E} \frac{d\sigma}
{\Vert\nabla H\Vert}\, \frac{M_1^\star}{\Vert\nabla H\Vert}\nonumber\\
& \simeq &\frac{(\delta E)}{N!}\langle\Vert\nabla H\Vert^{-1}\rangle 
\int_{\Sigma_E}\frac{d\sigma}{\Vert\nabla H\Vert}\ M_1^\star
\label{coarea2}
\end{eqnarray}
where, in the last approximate replacement,
we have used that $\Vert\nabla H\Vert$ is positive and only very weakly 
varying at large $N$.

By means of H\"older's inequality for integrals we get
\[
\int_{\Sigma_E}\frac{d\sigma}{\Vert\nabla H\Vert} M_1^\star \leq 
\left( \int_{\Sigma_E}\frac{d\sigma}{\Vert\nabla H\Vert} \vert M_1^\star\vert^N
\right)^\frac{1}{N}
\left( \int_{\Sigma_E}\frac{d\sigma}{\Vert\nabla H\Vert} \right)^\frac{N-1}{N}
~,
\]
the sign of equality being better approached when $M_1^\star$ is everywhere
positive.
Hence, using Eqs.~(\ref{volmu}), (\ref{coarea1}) and (\ref{coarea2}) 
\begin{equation}
\Omega_\nu (E) \leq [\Omega_\nu (E)]^\frac{N-1}{N}\left( \frac{1}{N!}
\int_{\Sigma_E}\frac{d\sigma}{\Vert\nabla H\Vert} \vert M_1^\star\vert^N
\right)^\frac{1}{N}\frac{\delta E}{\langle\Vert\nabla H\Vert\rangle}
\end{equation}
and introducing a suitable deformation factor $d(E)$ we can write
\begin{equation}
[\Omega_\nu (E)]^\frac{1}{N} = \frac{d(E)\delta E}
{\langle\Vert\nabla H\Vert\rangle}\left( \frac{1}{N!}
\int_{\Sigma_E}\frac{d\sigma}{\Vert\nabla H\Vert} \vert M_1^\star\vert^N
\right)^\frac{1}{N}
\end{equation}
so that 
\begin{equation}
\Omega_\nu (E) =  \frac{2^N[d(E)]^N(\delta E)^N}{\langle\Vert\nabla H\Vert
\rangle^{N+1}}\int_{\Sigma_E}d\sigma\ \left(\vert K_G\vert +\frac{{\cal R}(E)}
{2^N N!}\right)
\end{equation}
where we have used $2^{-N}\vert M_1^\star\vert^N=(\kappa_1+\kappa_2+\dots +
\kappa_N)^N= N!\vert K_G\vert + {\cal R}(E)$, with $\kappa_1,\dots,\kappa_N$
the principal curvatures of $\Sigma_E$, and $K_G$ is the Gauss-Kronecker
curvature of $\Sigma_E$, $K_G=\prod_{i=1}^N\kappa_i$. ${\cal R}(E)$ is a
remainder. Again we have used that $\Vert\nabla H\Vert$ is only very weakly 
varying at large $N$ and that it is always positive.

According to the Chern-Lashof theorem \cite{chernlashof}
\begin{equation}
\int_{\Sigma_E}d\sigma\ \vert K_G\vert = vol({\mathbb S}_1^{N-1})
\sum_{i=0}^N \mu_i(\Sigma_E)
\end{equation}
where ${\mathbb S}_1^{N-1}$ is an $N-1$-dimensional hypersphere of unit radius
and $\mu_i(\Sigma_E)$ are the Morse indexes of $\Sigma_E$ which are defined
exactly as those of the $M_v$'s seen in the previous Sections, i.e., as
the number $\mu$ of critical points of index $i$ on a given level set
$\Sigma_E=H^{-1}(E)$; a critical point is a point where $\nabla H=0$, the 
index $i$ of a critical point is the number of negative eigenvalues of the
Hessian of $H$ computed at the critical point. 

Finally the entropy per degree of freedom reads as 
\begin{eqnarray}
S(E)& =& \frac{1}{N} \log \Omega_\nu (E)\nonumber\\
&=&\frac{1}{N} \log \left[ vol({\mathbb S}_1^{N-1})\sum_{i=0}^N \mu_i(\Sigma_E)
+ \int_{\Sigma_E}d\sigma \frac{{\cal R}(E)}{2^N N!}\right]
+\frac{1}{N}  \log  \frac{2^N[d(E)]^N(\delta E)^N}{\langle\Vert\nabla H\Vert
\rangle^{N+1}}\ .
\label{entropy}
\end{eqnarray}
The meaning of Eq.~(\ref{entropy}) is better understood if we consider that the
Morse indexes $\mu_i(M)$ of a differentiable manifold $M$  are related to the 
Betti numbers $b_i(M)$ of the same manifold by the inequalities
\begin{equation}
\mu_i(M) \geq b_i(M)~.
\label{morseineq}
\end{equation}
The Betti numbers are fundamental {\it topological invariants} 
\cite{nakahara}
of differentiable manifolds; they are the diffeomorphism-invariant dimensions
of suitable vector spaces  (the de Rham's cohomology spaces), thus they are
integer numbers. The equality sign holds only for the so-called perfect Morse 
functions, which are rare, however, at large dimension we can safely replace
Eq.~(\ref{morseineq}) with $\mu_i(M) \simeq b_i(M)$ (see e.g.~Refs.~\cite{xy}).

Equation (\ref{entropy}), rewritten as
\begin{eqnarray}
S(E)&\simeq &
\frac{1}{N} \log \left[ vol({\mathbb S}_1^{N-1})\sum_{i=0}^N b_i(\Sigma_E)
+ \int_{\Sigma_E}d\sigma \frac{{\cal R}(E)}{2^N N!}\right]\nonumber \\ 
& +&\frac{1}{N}  \log  \frac{2^N[d(E)]^N(\delta E)^N}{\langle\Vert\nabla H\Vert
\rangle^{N+1}}\ ,
\label{entropia}
\end{eqnarray}
links topological properties of the {\it microscopic} phase space with
the {\it macroscopic} thermodynamic potential $S(E)$.

In particular, even though the function  ${\cal R}(E)$ is unknown, 
sudden changes of the topology of the hypersurfaces $\Sigma_E$ 
(reflected by the energy variation of $\sum b_i(\Sigma_E)$)  
necessarily affect the energy variation of the entropy.

Finally, we resort to the fact that -- at large $N$ -- the volume measure of
$\Sigma_E$ concentrates on ${\mathbb{S}}^{N-1}_{\langle 2K \rangle^{1/2}}
\times M_{\langle V\rangle}$, where ${\mathbb{S}}^{N-1}_{\langle 2K 
\rangle^{1/2}}=\{ (p_1,\dots,p_N)\vert \sum p_i^2=\langle 2K\rangle\}$ is the
kinetic energy hypersphere and
$M_{\langle V\rangle}=\{ (q_1,\dots,q_N)\vert V(q)\leq \langle V\rangle\}$,
 so that $\Sigma_E$ can be approximated by this 
product manifold,
and we resort to the Kunneth formula  \cite{nakahara} for the Betti numbers of
a product manifold $A\times B$, i.e.
\begin{equation}
b_i(A\times B) = \sum_{j+k=i} b_j(A) b_k(B)
\end{equation}
which, applied to ${\mathbb{S}}^{N-1}_{\langle 2K \rangle^{1/2}}\times 
M_{\langle V\rangle}$, gives $b_i(\Sigma_E)= 2 b_i(M_v)$ for 
$i = 1,\ldots,N-1$,
and $b_j(\Sigma_E)= b_j(M_v)$ for $j = 0,N$, since all the Betti numbers of an 
hypersphere vanish but $b_0$ and $b_N$ which are equal to $1$ \cite{nakahara}. 
Eventually we obtain
\begin{eqnarray}
S(v)&\simeq &
\frac{1}{N} \log \left[ vol({\mathbb S}_1^{N-1})\left( b_0 +
\sum_{i=1}^{N-1} 2 b_i(M_v ) + b_N \right)\right.\nonumber \\  
&+ &\left. {\tilde{\cal R}(E(v))}\right] 
 +\frac{1}{N}  \log  \frac{2^N[d(E)]^N(\delta E)^N}{\langle\Vert\nabla H\Vert
\rangle^{N+1}}\ ,
\label{entropiav}
\end{eqnarray}
where ${\tilde{\cal R}(E(v))}$ stands for the integral on the product manifold
of the remainder which appears in Eq.(\ref{entropia}). The equation above 
makes explicit the fact that the kinetic energy term of a standard Hamiltonian
is topologically trivial. 

From this equation we 
see that a fundamental topological quantity, the sum of the Betti
numbers of the submanifolds $M_v=\{ (q_1,\dots,q_N)\in{\Bbb R}^N\vert
V(q)\leq v\}$ of configuration space, is related, although with some 
approximation, to the thermodynamic entropy of the system.

\begin{figure}
\centerline{\psfig{file=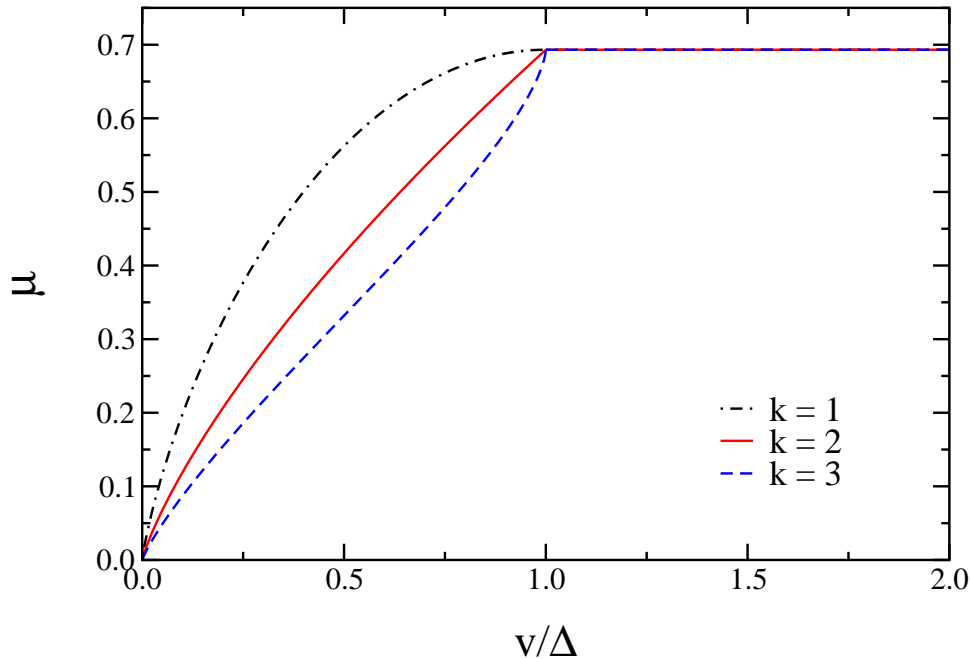,width=14cm,angle=-90}}
\caption{Logarithm of the sum of the Morse indexes divided by $N$,  
$\mu =\frac{1}{N}\log \left[ \mu_0 +2 \sum_{i=1}^{N-1} \mu_i(M_v )+\mu_N \right]$ 
of the manifolds $M_v$ versus the energy density $v$, scaled with
$\Delta$, for $k=1,2,3$. }
\label{sommamui}
\end{figure}

While a relationship between topology and thermodynamics exists, as is shown
by both Eqs.\ (\ref{entropiav}) and (\ref{entropy}), an analytic formula linking
the Euler characteristic to thermodynamics has not been found yet and is unlikely
to exist. Therefore, in those cases allowing the analytic computation of the Morse 
indexes (as for the models tackled in this paper), besides the 
computation of the Euler characteristic through the formula (\ref{defchi}), 
we can use the Morse indexes 
to replace the sum $\left( b_0 +\sum_{i=1}^{N-1} 2 b_i(M_v ) + b_N\right)$ in 
Eq.\ (\ref{entropiav}) with
$\left(  \mu_0 +\sum_{i=1}^{N-1} 2 \mu_i(M_v ) + \mu_N\right) $, having 
resorted again to the estimate $b_i(M)\simeq\mu_i(M)$. 
Then we can plot this sum, that we shall denote by $\mu$, against the potential
energy density for the three different cases considered: $k=1,2,3$.
The  result is shown in Fig.\ \ref{sommamui}, where sharp differences are evident
between the three situations: {\it i)} $k=1$, absence of any phase transition,
in which case the pattern of $\mu$ {\it vs} $v$ is smooth; {\it ii)} $k=2$,
second order phase transition, in which case the pattern of $\mu$ {\it vs} $v$
displays a sharp non-differentiable change at the phase transition point;
{\it iii)} $k=3$, first order phase transition, in which case the pattern 
of $\mu$ {\it vs} $v$ again displays a sharp non-differentiable change at the 
phase transition point which is approached from the left with an opposite 
concavity with respect to the second order transition case. Likewise the Euler 
characteristic,
the quantity $\mu$ is in one-to-one correspondence with topology changes of
the manifolds $M_v$, but $\mu$ has the advantage of being directly
related with thermodynamic entropy.

Both the general analytic result of Eq.\ (\ref{entropiav}) and the particular
analytic result obtained for the $k$TM and reported in Fig.\ \ref{sommamui}
are of great relevance in view of a deeper understanding of the relationship
between topology changes of configuration space and  
phase transitions: further work is ongoing along this direction.

As a final comment, let us remark that the clearcut differences among the
three different cases in  Fig.\ \ref{sommamui} are detected {\it prior to}
 and 
{\it independently of} the definition of any statistical measure in 
configuration space: the relevant information about the macroscopic physical 
behavior is already
contained in the microscopic interaction potential itself and concealed in its 
way of shaping configuration space submanifolds.

\section{Concluding remarks}

\label{conclusions}
We have presented a mean-field model whose thermodynamics is exactly solvable
in both the canonical and the microcanonical ensemble: the model depend on a
parameter $k$ and exhibits no
transitions, a continuous phase transition and a discountinuous one, in the
cases $k=1$, $k=2$, and $k \geq 3$, respectively. For this model, a clear
and sharp 
relation between the presence of a phase transition and a particular topology
change in the submanifolds of the $N$-dimensional configuration space is
analytically worked out: this correspondence becomes one-to-one if we look at
the submanifolds of a reduced two-dimensional configuration space spanned by
collective variables. Moreover, not only the presence and the location in energy
of the transition can be inferred by looking at the behavior of topological
quantities: also the order of the transition is related to the behavior of a
topological invariant of the above mentioned submanifolds, their Euler
characteristic. This suggests that topological quantities are linked in general
to thermodynamic observables.
Such a general link, although based on some approximations, has been derived in
the final Section of the paper. 

The results presented here confirm the validity and the potential of the
topological approach to phase transitions, which has recently received a
rigorous background via the proof of a theorem \cite{theorem} 
stating that, for systems with 
short-ranged interactions, topology changes in the submanifolds of configuration
space are a {\em necessary} condition for a phase transition to take place. 
The model we studied here is not short-ranged, thus the theorem might probably
be extended to a more general class of systems. However, we would like to
mention the case of a mean-field model, the fully coupled 
$\varphi^4$ model, which has been
recently studied \cite{phi4mf}, and where the relation between the topology
changes in the submanifolds of the configuration space and the phase transition
is less straightforward. Further work is then needed to assess
the full potential and the limits of the topological approach to phase
transitions.

\end{document}